\documentclass{aa}  

\usepackage{graphicx}
\usepackage{amsmath}
\usepackage{subfigure}
\usepackage{threeparttable}
\usepackage{txfonts}
\usepackage{graphicx,stackengine}
\newcommand\underlay[4]{%
  \stackengine{0pt}%
  {\kern#2\includegraphics[height=#1]{#4}}%
  {\includegraphics[height=#1]{#3}}%
  {O}{l}{F}{F}{L}%
}
\newcommand\addunderlay[4]{%
  \stackengine{0pt}%
  {\kern#2\includegraphics[height=#1]{#4}}%
  {#3}%
  {O}{l}{F}{F}{L}%
}

\begin{document}

   \title{Kinematical asymmetry in the dwarf irregular galaxy \object{WLM} and a perturbed halo potential}

   %\subtitle{I. Overviewing the $\kappa$-mechanism}

   \author{M. Khademi
          \inst{1,2}\fnmsep\thanks{ma\_khademi@sbu.ac.ir}
          \and
           Y. Yang  \inst{1}
          \and
          F. Hammer \inst{1}
          \and
          S. Nasiri  \inst{2}
          }

   \institute{GEPI, Observatoire de Paris, Universite PSL, CNRS, Place Jules Janssen 92195, Meudon, France\\
      \and
   Department of Physics, Shahid Beheshti University, G.C., Daneshjou Boulevard, District 1, 19839 Tehran,  Iran\\
            % \email{ yanbin.yang@obspm.fr, Francois.Hammer@obspm.fr}
                        }

 %  \date{Received September 15, 1996; accepted March 16, 1997}

  \abstract
  {\object{WLM} is a dwarf irregular that is seen almost edge-on that has prompted a number of kinematical studies investigating its rotation curve and its dark matter content.
In this paper, we investigate the origin of the strong asymmetry of the rotation curve, which shows a significant discrepancy between the approaching and the receding side. We first examine whether an $m = 1$ perturbation (lopsidedness) in the halo potential could be a mechanism creating such kinematical asymmetry. To do so, we fit a theoretical rotational velocity associated with an $m = 1$ perturbation in the halo potential model to the observed data via a $\chi-$squared minimization method. We show that a lopsided halo potential model can explain the asymmetry in the kinematic data reasonably well. We then verify that the kinematical classification of \object{WLM} shows that its velocity field is significantly perturbed due to both its asymmetrical rotation curve and also its peculiar velocity dispersion map.  In addition, based on a kinemetry analysis, we find that it is possible for \object{WLM} to lie in the transition region, where the disk and merger coexist. In conclusion, it appears that the rotation curve of \object{WLM} diverges significantly from that of an ideal rotating disk, which may significantly affect investigations of its dark matter content.}
  
   
 \maketitle             
%
%-------------------------------------------------------------------

\section{Introduction}
It has been found that disk galaxies have massive dark matter halos but little is known about the shape of such halos and their content. Dark halos were modeled as spherical until \cite{Binney1978} suggested that the natural shape of dark halos is triaxial. Triaxial dark matter halos come to pass naturally in cosmological simulations of structure formation in the universe. These simulations also illustrate that there may be a universal density profile for dark matter halos \citep{Nav1996, Col1996}, but the accurate distribution of halo shapes is still unclear. The observed distribution of the shapes of dark matter halos can be a constraint on the scenarios of galaxy formation and evolution. Investigating the shapes of dark halos can be done in two parts: measuring the ratio $ c/a $ , namely, flattening perpendicular to the plane of the disk and measuring the intermediate to major axis ratio $ b/a $ , namely, the elongation of the potential, which is of type $m = 2$ perturbation in the potential and useful to measure the elongation of orbits. The effects of a global elongation of the dark matter halo are similar to an $m = 2$ spiral arm \citep{sch}.

Asymmetries in the distribution of light and the neutral hydrogen gas $HI$ are often observed in spiral galaxies and it has been known for a long time that the light distribution and therefore the mass distribution in disks of spiral galaxies is not closely axisymmetric; for example isophotes of  $M101$ and $NGC 1637$  are elongated in one half of these galaxies \citep{san}. Such an asymmetry,  which was detected for the first time in the spatial extent of the atomic hydrogen gas in the outer regions in two halves of some galaxies and also in the distribution of light \citep{bal}. A galaxy showing a global non-axisymmetric spatial distribution of type $m = 1$ perturbation in the potential or a $cos (\phi)$ distribution, where $\phi$ is the azimuthal angle in the plane of the disk, is a lopsided galaxy \citep{jog}. The asymmetric distribution of  $HI$ gas ( morphological lopsidedness in $HI$ gas ) in spiral galaxies such as $M101$ was reported in the early $HI$ observations \citep{Beal}. These kinds of galaxies, which are more extended on one side than the other,  were named lopsided galaxies.

Morphological lopsidedness was confirmed in a Fourier-analysis study of a large sample of 149 galaxies, with about one third of galaxies of this sample showing asymmetry in the amplitude of the $m = 1$ Fourier component \citep{bou}. Thus, morphological lopsidedness in the disk is a general phenomenon. So, it is important to investigate the origin and dynamics of the lopsided distribution in the galaxies. 
The lopsided (perturbed) distribution in the atomic hydrogen gas has been mapped morphologically \citep{hay} as well as kinematically for a few galaxies \citep{sch,Swa1999} and by global velocity profiles for a larger sample \citep{ric}. Also, such an asymmetry has been found in dwarf galaxies \citep{swa2} and in the star-forming regions in the irregular galaxies \citep{heller}. The asymmetry may affect all scales in a galaxy, but the large-scale lopsidedness is more conspicuous. 
This kind of $m = 1$ perturbation is expected to have a significant impact on the dynamics of the galaxies, their evolution, and the star formation within them; in addition, it  may also play an important role in the growth of the central black hole and on the nuclear fueling of the active galactic nucleus (AGN) in a galaxy \citep{JogJog,jog2009}.
A perturbation in the gravitational potential of a dark matter halo can make a galaxy asymmetric and create asymmetry in  the gas surface density profile \citep{jog,Jog2000}. Also, a symmetric dark matter halo can produce an asymmetric galaxy if the disk orbits of-center with respect to the overall potential \citep{Lev}.

A new area that is beginning to be investigated is the asymmetry at the centers of mergers of galaxies. A systematic study was recently carried out by \cite{Jog2006}, in order to understand the lopsidedness of the intensity and luminosity distribution and hence the mass asymmetry within their central few $kpc$ regions of mergers.
Recently \cite{Ghosh2021} did a simulation study of lopsided asymmetry generated in a minor merger to investigate the dynamical effect of the minor merger of galaxies on the excitation of lopsidedness.
The lopsided modes and the central asymmetry that is merger-driven can play an important role in the dynamical evolution of the central regions, especially with regard to the star formation taking place within it, and on the nuclear fueling by outward transporting of the angular momentum. Therefore, these process can be important in the hierarchical evolution of galaxies \citep{jog2009}.

Kinematical lopsidedness or a $\cos(\phi)$ asymmetry is often also observed in the kinematics of the galaxies and therefore in the velocity fields \citep{sch} and in the rotation curves on two halves of a galactic disk \citep{Swa1999,Sof}.
A galaxy showing a spatial asymmetry between two sides of the galaxy would naturally show kinematical asymmetry except in the cases of face-on galaxies. A face-on galaxy may show morphological asymmetry but not kinematical asymmetry \citep{Jog2000,jog2002,jog2009}. However, in the past, a number of authors have made a distinction between the spatial or morphological lopsidedness and kinematical lopsidedness \citep{Swa1999,noor} and have even claimed \citep{korn} that the velocity asymmetry is not always correlated with the morphological asymmetry. 

Dwarf irregulars are the most common type of galaxies in the local universe (see, e.g., \cite{Dale}. Such galaxies with extended $HI$ disk distributions allow measurement of rotation curves and hence investigate dark matter halo properties and its shape to large radial distances, beyond the optical disk. These types of galaxies contain a huge reservoir of dark matter halo that dominates over most of disk \citep{GHOSH201838}.

\object{WLM} is a near edge-on gas-rich dwarf irregular galaxy in the Local Group. This galaxy is rotationally supported and clearly rotating and isolated from massive galaxies \citep{kep}. 
It is worth noting that \object{WLM} in the local group, as well as most of those dwarf irregular galaxies identified as kinematically lopsided, is an isolated system and does not show any signs of strong tidal interaction and there is currently no clear evidence for ongoing accretion of satellite galaxies. The dwarf irregular galaxy \object{WLM} exhibits interesting dynamics, which was recently studied in a number of works. 
Its rotation curve is asymmetric, the rotation curves for the approaching and the receding halves of \object{WLM} are not symmetric and they are distinctly different. In this galaxy, the rotation curve of the receding side rises much more slowly than for the approaching side, while the velocity gradient for the approaching side is steeper than for the receding side in the outer region for this galaxy. The rotation curve of the approaching side rises and then flattens at a certain radius, whereas the rotation curve for the receding side continues to rise \citep{kep}.
\object{WLM}'s velocity field map obtained from an observational $HI$ data cube is asymmetric too. The spider-like shape of iso-velocity contours are curved more strongly on the approaching side and take on the characteristic shape of differential rotation, while the iso-velocity contours on the receding side remain more or less straight and mostly parallel to the minor axis, which is consistent with regular solid-body rotation.
Dwarf irregular galaxies usually have a slowly rising rotation curve close to a solid body and there is lack of differential rotation in dwarf irregular galaxies \citep{kep,Lelli}. 
Also, the global HI emission profile of \object{WLM} shows a strong asymmetry \citep{kep,Rog2020}. 
Global $HI$ line profiles are influenced by both of the kinematics and the $HI$ distribution. The shape of the global line profile has been determined by both the kinematics and the density distribution of the gas. Galaxies with such asymmetric global line profiles often have rotation curves that are steeper on one side of the galaxy than on the other side \citep{Swa1999}.

This paper is organized as follows. In Sect.~\ref{Velocity Fields Models}, we derive the velocity fields map from two different kinematic models for this galaxy, solid body rotation, and differential rotation, if it is shown that this galaxy orbits in the potential of an axisymmetric dark halo. In Sect. \ref{DYNAMICS OF ORBITS}, we review the dynamics of orbits in a lopsided halo potential (non-axisymmetric potential) and extract the rotation curve of \object{WLM} by fitting a theoretical rotational velocity associated with an $m = 1$ perturbation in the halo potential model to the observed data for both of receding and approaching sides.
We also generate a velocity field map and surface density map for the $HI$ gas disk and stellar disk from this model, associated with an $m = 1$ perturbation in the halo potential (a lopsided halo potential model).
In Sect. \ref{Classification}, based on kinematic asymmetries, for the purpose of investigating whether a merger could have created such a lopsidedness in the halo potential, we determine the amplitude of velocity asymmetry and the strength of deviation of the velocity field of a perturbed rotation model (an $m=1$ perturbation in the potential) from that of an ideal rotating disk case by obtaining the relative level of deviation of the velocity field from that of an ideal rotating disk case.  In addition, based on $\sigma$ - centering, we classify this galaxy as having perturbed rotations. Finally, in Sect. \ref{Conclusion}, we present our concluding remarks.

\section{Modeling the velocity fields of the rotating disks and different types of galactic rotation}\label{Velocity Fields Models}
In a rotating disk, the bulk motions can be projected as ellipses on the sky.  The orientation of such ellipses depends on the disk inclination and kinematic position-angle. The observed line-of-sight (LOS) radial velocity, $ v_{los} $, extracted along such ellipses can be expressed as:\   
\begin{align}
v_{los} (R, \psi) = &v_{c}(R, \psi)\sin(i)\cos(\psi) + v_{r}(R, \psi)\sin(i)\sin(\psi) + \nonumber\\
&v_{z} (R, \psi) \cos(i) + v_{sys.} 
\end{align}
Here, $v_{sys}$ is the systemic velocity of the galaxy that corresponds to the galaxy red-shift, $ v_{c}(R) $ is the circular (rotational) velocity in the azimuthal $\psi$ direction, $v_{r}$ is the radial velocity in the disk plane, and $v_{z} $ traces the vertical motions. The azimuthal angle $\psi$  is measured from the projected major axis in the plane of galaxy,  $ R $ is the radius of a circular ring in that same plane (or the semi-major axis length of the ellipse once projected on sky), and $ i $ is the inclination of the disk ($ i = 0 $ for a disk seen face-on) \citep{2017bookHammer}.
\subsection{Velocity fields of axisymmetric disk galaxies \\( 2D model: Rotating disk)}
If the cold HI gas in a disk galaxy orbits in the potential of an axisymmetric dark halo (a static, spherically symmetric gravitational potential field), the resulting density field and velocity field are axisymmetric as well. This means that the velocity field will only show pure circular rotation and the predominant motion of gas in a spiral galaxy is rotation and the observed velocity fields of disk galaxies generally can be fitted perfectly by circular motion.
Assuming that at radius R, a gas cloud follows a near-circular path with speed $ v_{c}(R)$. All we can detect of this motion is the LOS radial velocity, $ v_{los}$, toward or away from us; its value at the galaxy's centre, $v_{sys}$, is the systemic velocity. For an ideal rotating disk  (a disk with pure rotational motion), the observed velocity fields are fitted with circular velocity and with no radial or out of plane motions (noncircular motions). In such an ideal rotating thin disk, the bulk motions draw circular orbits in the plane of galaxy, projected on-sky as ellipses \citep{Krajnovic2006}.
For a given projected elliptical HI ring, the projected velocity along the LOS (radial velocity) is:
\begin{equation}\label{difeq}
v_{los} (R, \psi) = v_{c}(R) \sin(i) \cos(\psi) + v_{sys}
,\end{equation}
where $v_{sys}$ is the systemic velocity of the galaxy, $ v_{c}(R) $ is the circular (rotational) velocity of the gas at radius $ R $, $\psi$ is the azimuthal angle of the rings in the plane of the galaxy (giving a star's position in its orbit), and $ i $ is the disk inclination ($ i = 0 $ for a disk seen face-on). The observed LOS radial velocity, $ v_{los} $, toward or away from an observer, is usually measured using the Doppler shift of emission line or absorption line in the spectra of $HI$ gas or stars.\\ 
\subsection{Solid-body rotation}
A homogeneous density distribution produces solid-body rotation with $\Omega=v_{c}(R)/R=constant$. When we plug this into Eq. (\ref{difeq}), we obtain:\ \begin{equation}
v_{los}  = \Omega R \sin(i) \cos(\psi) + v_{sys}
.\end{equation}
Since $x=R \cos(\psi)$, this gives: 
\begin{equation}
v_{los}  = \Omega x \sin(i) + v_{sys}
.\end{equation}
Therefore the velocity does not depend on $y$ and contours of constant velocity will be parallel to the $y$ axis. Vertical contours of the constant velocity near the centre of a two-dimensional velocity field is indicative of the solid-body rotation.
\subsection{Differential rotation}
\begin{figure*} 
\subfigure[]{ \includegraphics[width=\columnwidth]{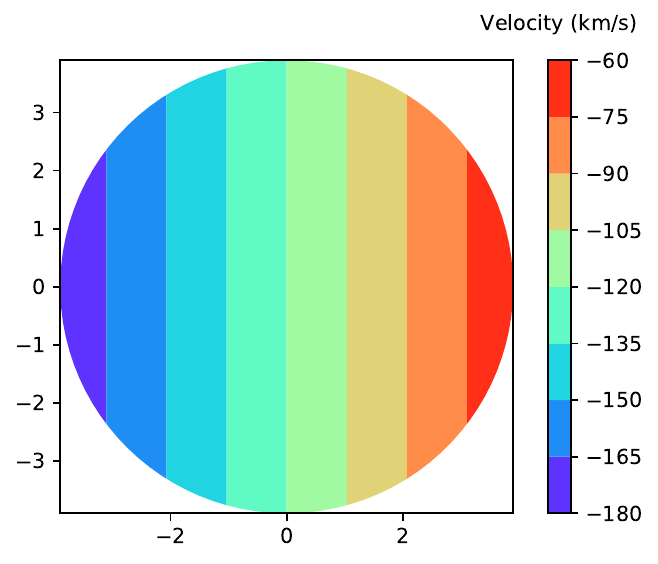} }\quad
   \subfigure[]{ \includegraphics[width=\columnwidth]{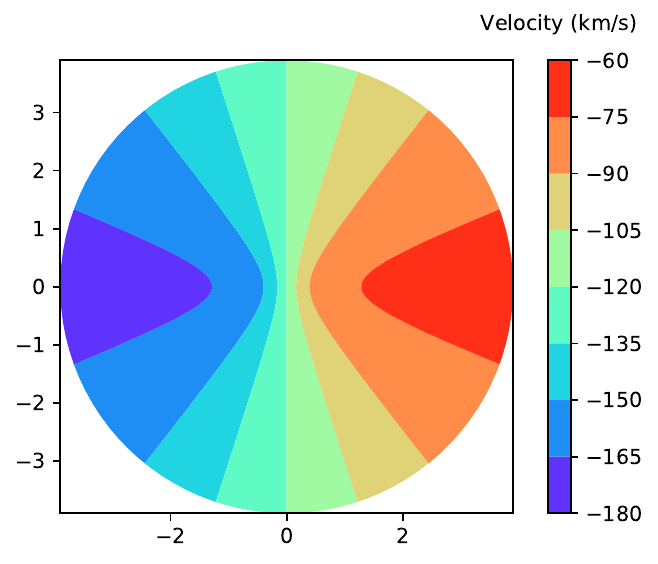} }
    \caption{ Pure circular rotations: (a) Velocity field map for the solid body rotation model, obtained from the model cube. Iso-velocity contours are mostly parallel to the minor axis, indicating a solid-body rotation. (b) Velocity field map for a differentially rotating disk model (rising-to-flat), obtained from the model-cube. Iso-velocity contours of the velocity field map display strong curvatures indicating of differential rotation.}
\label{VFM}
\end{figure*}

A simple model for differential rotation is cored logarithmic potential that can produce a rising rotation curve that flattens at a certain radius (rising-to-flat rotation curve):
\begin{equation}
V_{0}(R) = \dfrac{1}{2}v_{0}^{2} \ln[ R_{c}^{2}+ R^{2}] 
,\end{equation}
with
\begin{equation}
v_{c}\left( R\right) = \dfrac{v_{0}R}{\sqrt{R^{2} + R_{c}^{2}}}
,\end{equation}
where $R_{c}$ and $ v_{0} $ are constants that $ v_{0} $  is circular velocity at large $ R $.
At $R \ll R_{c}$ , this corresponds to solid-body rotation and at $R \gg R_{c}$ the rotation curve is flat; this model is therefore a good representation of a rising-to-flat rotation.\\
By programming in Python, we generated the velocity field maps from two different kinematic models for this galaxy: solid body rotation (slowly rising rotation curve) and differential rotation (steeply rising-to-flat rotation curve), as shown in Fig. \ref{VFM}. This figures show contours of constant velocities and the major axis of these velocity field map is horizontal for each of these models. \\
\begin{table*}
\caption{Photometric and structural parameters of \object{WLM}.}
\label{tab1}
 \begin{threeparttable}
\begin{tabular}{cccccccccc}
\hline
(1)&(2)&(3)&(4)&(5)&(6)&(7)&(8)&(9)&(10)\\
  position angle& Median inclination &Distance & $V_{sys}(km/s)$& $\epsilon=1-\dfrac{b}{a}$ &$R_{h}(pc)$&$R_{max}(pc)$&$M_{v}(mag)$& $f_{c}$ & $F_{e}/H$\\
(degree)& angle (degree)& (kpc)&&&&&&&\\
 \hline
  4\tnote{a} &  75\tnote{b}& 933\tnote{a} &  $-122 \pm 4 $ \tnote{c} &0.65\tnote{a} & 2111\tnote{a} & 2304 \tnote{b} & -14.2\tnote{a} & 4.8\tnote{d} &-1.27\tnote{a} \\
  \hline
 \end{tabular}
   \begin{tablenotes}
   \item\label{1} References: (a) \citep{Mc}, (b) \citep{kep} , (c) \citep{Rog2020} and (d) \citep{Io}
 \end{tablenotes}
  \end{threeparttable}
\end{table*}
In Table \ref{tab1}, col. 1 gives position angle of major axis, measured in degree, col. 2 gives the median inclination angle of galaxy in unit of degree, and col. 3 gives heliocentric distance. Column .4 gives systematic velocity of galaxy, col. 5 gives the ellipticity, $ \epsilon = 1 - b/a $, where a and and b are the semi-major axis length and semi-minor axis length, respectively. The ratio of the semi-major axis $a$ and the semi-minor axis $b$ quantifies how far the isophote (contour of constant surface brightness) differs from a circle. Column. 6 gives half-light radius in parsec and Col. 7 gives radius of extension of HI gas in parsec. Column. 8 gives absolute visual magnitude, Col. 9 gives conversion factor from arcsec to pc, and Col. 10 gives mean stellar metallicity.\\
\section{Dynamics of orbits in a lopsided potential (description of a perturbed velocity field model)}\label{DYNAMICS OF ORBITS}
\begin{figure}
	\includegraphics[width=\columnwidth]{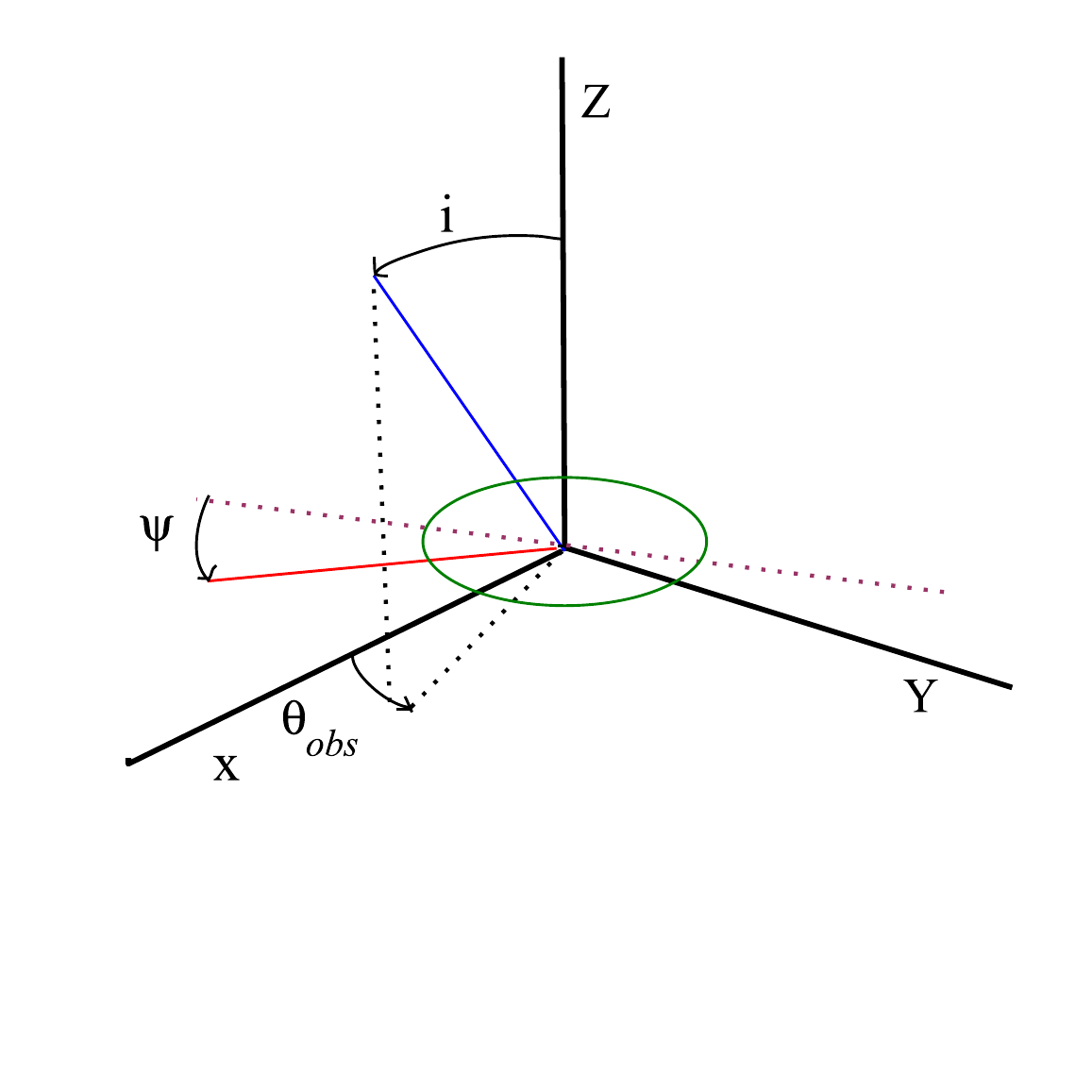}
   \vspace{-2.2cm}
    \caption{Geometry of the projected orbit which is indicated by green ellipse and lies in X-Y plane. The orbit is elongated along the Y axis. The Z$-$axis is perpendicular to X-Y plane, i.e., the plane of the orbit. The viewing angle $ \theta_{obs}$ and $ i $ are the polar coordinate of the line-of sight. The inclination angle $i$ is measured from the Z$-$axis. The angle $\theta_{obs}$ is measured from the line $ \theta = 0$. The azimuthal angle $\psi$ is the angle in the plane of the orbit and it's defined to be zero at the line of nodes.}
    \label{Fig.Geometery}
\end{figure}

A simple rotating disk cannot reproduce the asymmetries that are seen in the observed velocity field and in the rotation curve of \object{WLM}.
The asymmetry in the rotation curve between approaching and receding sides of the galaxy, rising more steeply on one side than on the other side, is a signature of the kinematical lopsidedness in the galaxy \citep{Swa1999}. We are going to investigate whether the kinematical asymmetry in \object{WLM} may be related to lopsidedness in the halo potential and kinematic lopsidedness can be caused by a perturbed halo potential and whether this galaxy is lopsided in its kinematics.\\
Assuming that the galaxy is a rotating disk and then investigating the model deviations to verify the accuracy of this assumption is a simple way to study kinematics of a galaxy \citep{2017bookHammer}.\\ 
Here, we analyse the case of the small deviations from axisymmetry of the potential $V$ of a filled gaseous disk, which can be written as a sum of harmonic components. The assumption that the potential includes a small perturbation and that the gas moves on the stable closed orbits allow us to use epicycle theory to analyse the velocity fields that are caused by such a perturbed potential.\\
The orbits are solved via first-order epicyclic theory. The treatment of orbits in the weak non-axisymmetric potentials is closely related to the epicycle theory of nearly circular orbits in an axisymmetric potential. To aid in the description of the non-axisymmetric features of the motion of the gas, the velocity fields have been decomposed into harmonic components along individual elliptical rings following the approach presented by \cite{sch}.
For studying the dynamics of particles ($HI $gas) in a galactic disk perturbed by a lopsided dark matter halo potential, we use the equations of motion in this perturbed potential for the possible closed orbits. In this way, the observed velocity field can be directly related to the potential of disk galaxies.
Here, we analyze the case of a small perturbation in the potential $V$, which can be written as a sum of harmonic components. \cite{sch} presented a method for measuring small deviations from symmetry in the velocity field of a filled gas disk which arises from small perturbation in the potential. This method is based on a higher-order harmonic expansion of the velocity field of the disk. This expansion was done by first fitting a tilted-ring model to the velocity field of the gaseous disk and subsequently expanding the full velocity field along each ring into its harmonic terms. The epicycle theory has been used to derive equations for the harmonic terms in a perturbed potential \citep{sch}.\\
{\bf Harmonic expansion}\\
The symmetries in kinematic of galaxies can be measured via the kinemetry method developed and explained by \cite{Krajnovic2006} and \cite{Shapiro2008}. A velocity field map can be divided into a number of elliptical rings and along each elliptical ring, the moment as a function of angle is extracted and decomposed into the Fourier series:
\begin{equation} 
v_{los} (\psi) = c_{0} + \sum\limits_{n = 1} c_{n}\cos(n\psi)+s_{n}\sin(n \psi) 
,\end{equation}
where $c_{0} $, gives the systemic velocity of each elliptical ring. This allows the velocity profiles to be described by a finite number of harmonic terms as well. So, we can express the LOS velocity as
\begin{equation} 
v_{los} (R, \psi) = v_{sys} + \sum\limits_{n = 1} c_{n}\cos(n\psi)+s_{n}\sin(n \psi)
.\end{equation}
We can calculate the amplitude $k_{n}$ of each Fourier harmonic order $n$ from the $c_{n}$ and $s_{n}$ coefficients:
\begin{equation} 
k_{n} = \sqrt{c_{n}^{2} + s_{n}^{2}}
.\end{equation}
The LOS velocity field in an ideal rotating disk (in pure circular rotation) is dominated by the $\cos(\psi)$ term. The velocity field of a disk that orbits in a non-perturbed potential shows only pure circular rotation that is given by:
\begin{equation}  
v_{los} = v_{c}\left(R \right) \sin\left( i\right) \cos\left( \psi\right). 
\end{equation}\\
We project the velocity field on the sky. The LOS velocity field is given by:
\begin{align}
v_{los}(R)=&\left[v_{R}\cos(\theta-\theta_{obs})-v_{\phi}\sin(\theta-\theta_{obs})\right]\sin(i),\nonumber\\
v_{los}=&\left[v_{R}\cos(\phi-\phi_{obs})-v_{\phi}\sin(\phi-\phi_{obs})\right]\sin(i).
\end{align}
This velocity field is observed from a direction $(\theta_{obs}, i)$, where the angle $i$ is the inclination of the plane of the disk with respect to the observer and $R$ and $\theta$ are polar coordinates in the rest frame (non-rotating frame) of the galaxy and $\theta_{obs}$ is  is defined as the angle between the line $ \theta = 0 $ and the observer (see Fig. \ref{Fig.Geometery}), but $\phi_{obs}$ is the viewing angle,  the angle in the rotating frame that corresponds to $\theta_{obs}$ (the angle between the line $\phi= 0$ and the observer that is zero along the major axis of the orbits). 

By introducing the variable $\psi=\theta-\theta_{obs}+\pi/2=\phi-\phi_{obs}+\pi/2$, we then have:
\begin{equation}
v_{los}=v_{R}\sin(\psi) \sin (i) + v_{\phi}\cos(\psi) \sin (i)
.\end{equation}
The angle $\psi$ is measured along the orbit, this angle is zero at the line of nodes. Sometimes this angle called the azimuthal angle and is defined by the orientation of the galaxy on the sky and is independent of the internal coordinate system.\\
By replacing $\phi\rightarrow\psi +\phi_{obs}-\pi/2$ in the expressions for $v_{R}$ and $v_{\phi}$, expanding the LOS velocity in multiple angles of $\psi$ and defining:
\begin{equation}
v_{\ast}=v_{c}(R)\sin i
,\end{equation}
and also assuming to the first order $\phi_{0} \approx\phi$, the LOS velocity field has the following form (\citealt{sch} and \citealt{ShoenmakerThesis}):
\begin{align}
v_{los}=&\sum\limits_{n}c_{n}\cos(n\psi)+s_{n}\sin(n \psi),\\
v_{los}=&c_{1}\cos \psi +s_{m-1}\sin(m-1)\psi + c_{m-1}\cos(m-1)\psi +\nonumber\\
& s_{m+1}\sin(m+1)\psi +c_{m+1}\cos(m+1)\psi,
\end{align}
where
\begin{align}\label{eq820}
c_{1}=&v_{\ast},\nonumber\\
s_{m-1}=&v_{\ast}(-\dfrac{1}{4}\left\lbrace [m-(m+1)\omega_{m}+\alpha]a_{1m}+(1-\omega_{m})a_{3m}\right\rbrace \sin(m\varphi_{m})\nonumber\\
&+\dfrac{1}{2}\lbrace m(1-\omega_{m})a_{4m}+[m(1-\omega_{m})-1+\alpha]a_{2m}\rbrace\cos(m\varphi_{m})),\nonumber\\
s_{m+1}=&v_{\ast}(\dfrac{1}{4}\left\lbrace [m-(m-1)\omega_{m}-\alpha]a_{1m}-(1-\omega_{m})a_{3m}\right\rbrace \sin(m\varphi_{m})\nonumber\\
&+\dfrac{1}{2}\lbrace m(1-\omega_{m})a_{4m}- [m(1-\omega_{m})+1-\alpha]a_{2m}\rbrace\cos(m\varphi_{m})),\nonumber\\
c_{m+1}=&v_{\ast}(-\dfrac{1}{4}\left\lbrace [m-(m-1)\omega_{m}-\alpha]a_{1m}-(1-\omega_{m})a_{3m}\right\rbrace \cos(m\varphi_{m})\nonumber\\
&+\dfrac{1}{2}\lbrace m(1-\omega_{m})a_{4m}-[m(1-\omega_{m})+1-\alpha]a_{2m}\rbrace\sin(m\varphi_{m})),
\end{align}

where $v_{\ast}=v_{c}(R)\sin i $, $\alpha = {d \ln[v_{c}(R)]}/{d \ln(R)}$ and $\varphi_{m} = \phi_{obs}-\pi/2-\phi_{m}(R)$. The explicit computations and derivation of the kinematic asymmetry based on the harmonic expansion have been presented by \cite{sch} and \cite{ShoenmakerThesis}. From Eq. (\ref{eq820}), we can find whether the potential includes a perturbation of harmonic number $m$, the LOS velocity field includes the $m + 1$ and $ m - 1$ terms. Thus, in the case of a harmonic number $ m = 1$ term in the potential (causing morphological lopsidedness), the LOS velocity field will include an $m = 0 $ term and an $m= 2$ term.
\subsection{Model velocity field map, rotation curve and surface density in a lopsided potential}
In this section, we generate a velocity field map, rotation curve, and surface density map for a perturbed potential to investigate whether such a potential is capable of creating asymmetry in the kinematics and the distribution of the galaxy. 
\subsubsection{Effect of $m=1$ distortion and making some simplifying assumptions}\label{simplifying}
We can map the velocity field in a lopsided potential from the measured harmonic terms. For this aim we make some simplifying assumptions: 1) the $m=1$ potential perturbation is dominant over the $m=3$ term; 2) the pattern speed of the $m=1$ perturbation is zero: $\omega_{1}=0$; 3) there is no radial dependence of the phase of perturbation: $\phi_{1}(R)=const$  $ \Rightarrow $ $ \phi^{'}_{1}(R)=0 $.
%\begin{enumerate}
%\item  The $m=1$ potential perturbation is dominant over the $m=3$ term.
%\item The pattern speed of the $m=1$ perturbation is zero: $\omega_{1}=0$.
%\item  There is no radial dependence of the phase of perturbation: $\phi_{1}(R)=const$  $ \Rightarrow $ $ \phi^{'}_{1}(R)=0 $.
%\end{enumerate}
So the LOS velocity field can be obtained as follows (the explicit computations were previously derived by \citealt{sch} and \citealt{ShoenmakerThesis}) :
\begin{equation}
v_{los}=\sum\limits_{n}c_{n}\cos(n \psi)+s_{n}\sin (n \psi)
,\end{equation}
If the potential includes a perturbation of harmonic number $m$, the LOS velocity field includes $m + 1$ and $ m - 1$ terms:\begin{align}
v_{los}=&c_{1}\cos\psi + s_{m-1}\sin(m-1)\psi +c_{m-1}\cos(m-1)\psi+\nonumber\\
&s_{m+1}\sin(m+1)\psi +c_{m+1}\cos(m+1)\psi.
\end{align}
In the case of a harmonic number $ m = 1$ term in the potential, the LOS velocity field will include an $m = 0 $ term and an $m= 2$ term. So, for  $m=1$, we have:
\begin{equation}
v_{los}=c_{1}\cos \psi +c_{0}+s_{2}\sin(2\psi)+c_{2}\cos(2\psi),
\end{equation}
with:
\begin{align}
c_{0}=&\dfrac{v_{c}\sin i}{2v_{c}(1+2\alpha) }[3V_{1}(R)+(2+\alpha)RV_{1}^{'}(R)]\cos \varphi_{1},\\
c_{2}=&-\dfrac{\sin i}{2v_{c}(1+2\alpha)}[V_{1}(R)-\alpha R V_{1}^{'}(R)]\cos \varphi_{1},\\
s_{2}=&\dfrac{\sin i}{2v_{c}(1+2\alpha)}[V_{1}(R)-\alpha R V_{1}^{'}(R)]\sin \varphi_{1},\\
c_{1}=&v_{c}\sin(i).
\end{align}
with $i$ as the inclination angle of the galaxy and $\varphi_{1}=\phi_{obs}-\pi/2 = \arctan\left(-\dfrac{s_{2}}{c_{2}}\right) $. Here, $\phi_{1}$ is one of the viewing angles, the angle in the plane of the orbit between the minor axis of the elongated orbit, and the observer; $\phi_{obs}$ is the viewing angle of the external observer, the angle between the line $\phi = 0$ and the observer that is zero along the major axis of the orbits. The explicit equations were derived by \citealt{sch} and \citealt{ShoenmakerThesis}.

{\bf The cored logarithmic potentials}\\
A simple model set-up for a non-perturbed potential to cause a velocity pattern that corresponds to differential rotation (rising to flat rotation, a rising rotation curve that tend to be flat at large radius) is a cored logarithmic potential:
\begin{equation}
V_{0}(R) = \dfrac{1}{2}v_{0}^{2} \ln[ R_{c}^{2}+ R^{2}] 
,\end{equation}
which has
\begin{equation}
v_{c}\left( R\right) = \dfrac{v_{0}R}{\sqrt{R^{2} + R_{c}^{2}}}
,\end{equation}
where $R_{c}$ and $ v_{0} $ are constants that $ v_{0} $  is circular velocity at large radius $ R $. 
At $R \ll R_{c}$, this equation corresponds to a solid-body rotation and at $R \gg R_{c} $ the rotation curve is flat. Thus this model is a good representation of a rising-to-flat rotation curve (differential rotation). \\
The orbit in a non-rotating logarithmic potential (perturbed potential, planar non-axisymmetric potentials), which is just the potential of the harmonic oscillator is the sum of independent harmonic motions (\citealt{Binney1978, binbook}).\\
For a potential that is perturbed by an $m = 1$ distortion and is changed by first-order perturbation, we can take the net potential, $V,$ at a given radius, $R,$ to be a sum of the non-perturbed potential, $V_{0}$, and the first harmonic component of the perturbation, $V_{1}(R)\cos(\phi)$, therefore, the total potential is:
\begin{equation}
V(R,\phi)=V_{0} (R)+V_{1}(R)\cos\phi,
\end{equation}
where chose the perturbation $V_{1}(R)$ to take the form \citep{jog}: 
\begin{equation}
V_{1}(R)=\epsilon_{lop}(R)v_{e}^{2},
\end{equation}
where $\epsilon_{lop}$ is a small perturbation parameter and $ v_{e} $ is the velocity of the flat part of the rotation curve.
Thus, the net potential is
 \begin{equation}
V(R,\phi)=V_{0} (R)+\epsilon_{lop}(R)v_{e}^{2}\cos\phi,
\end{equation}
with this choice for $V_{1}(R)$ we have
\begin{equation}
a_{11}=\dfrac{4\epsilon_{lop}}{v_{t}^{2}(1+2\alpha)},
\end{equation}
\begin{equation}
a_{31}=\dfrac{2(1-2\alpha)\epsilon_{lop}}{v_{t}^{2}(1+2\alpha)}.
\end{equation}
%For a cored logarithmic potential, $\alpha= {v_{0}R_{c}^{2}}/({R^{2} + R_{c}^{2}})$.\\
The orbits can be found as follows
\begin{equation}
R=\dfrac{R_{0}}{v_{t}^2}\left(1-2\epsilon_{lop}(R)\cos \phi_{0}\right)
,\end{equation}
\begin{equation}
\phi=\phi_{0}+\dfrac{(3-2\alpha)\epsilon_{lop}(R)\sin\phi_{0}}{v_{t}^{2}(1-2\alpha)},
\end{equation}
where
\begin{equation}
v_{t}(R)=\dfrac{v_{c}(R)}{v_{e}},
\end{equation}
and the velocities can be given as
\begin{align}
v_{R}=&\dfrac{2v_{c}(R)}{v_{t}^{2}}\epsilon_{lop}(R)\sin(\phi_{0}),\\
v_{\phi}=&v_{c}(R)\left[1+\dfrac{\epsilon_{lop}(R)}{v_{t}^{2}}\cos\phi_{0}\right],
\end{align}
The LOS velocity field is
\begin{align}
v_{los}=&\sum\limits_{n}c_{n}\cos(n\psi)+s_{n}\sin(n \psi),\\
v_{los}=&[v_{R}\sin\psi + v_{\varphi}\cos \psi]\sin i,
\end{align}
and 
\begin{equation}\label{eqvlos}
v_{los}=c_{1}\cos \psi +c_{0}+s_{2}\sin(2\psi)+c_{2}\cos(2\psi)
.\end{equation}
Finally, we can obtain the measured harmonics as follows:
\begin{align}
c_{0}=&\dfrac{v_{c}}{v_{t}^2}\sin i\left(\dfrac{3}{2(1+\alpha(R))}\right)\epsilon_{lop}(R)\cos\phi_{1},\label{eqc0}\\
c_{2}=&-\dfrac{v_{c}}{v_{t}^2}\sin i\left(\dfrac{1}{2(1+\alpha(R))}\right)\epsilon_{lop}(R)\cos\phi_{1},\label{eqc2}\\
s_{2}=&\dfrac{v_{c}}{v_{t}^2}\sin i\left(\dfrac{1}{2(1+\alpha(R))}\right)\epsilon_{lop}(R)\sin\phi_{1},\label{eqs2}\\
c_{1}=&v_{c}\sin(i)\label{eqc1}.
\end{align}
We note that the LOS velocity field of a pure rotational disk (a disk in pure circular rotation) is $ v_{los}=v_{c}(R)\sin i \cos \psi $.\\

For an $m = 1$ perturbation in the halo potential model, the LOS velocity field is presented by Eqs. (\ref{eqvlos})$-$(\ref{eqc1}). Our analysis is based on a reduced$-\chi^{2}$ as the goodness-of-fit statistic. We fit the theoretical rotational velocity extracted from the theoretical LOS velocity field, which is a function of three free parameters: $\epsilon_{lop}$, $v_{e} $, and $\phi_{1}$ to the observed data points via an $\chi-$squared minimization method (least squares fitting) for both of the approaching and receding sides by programming in Python. 
\begin{equation}
v_{los} = f (R, \epsilon_{lop}, v_{e}, \phi_{1}).
\end{equation}

The best-fit value of each parameter, $\epsilon_{lop}$, $v_{e}$, and $\phi_{1}$ in this model for the best fit to the observed rotation curve are obtained by minimizing this function:
\begin{equation}\label{chy}
\chi_{r}^{2}=\dfrac{1}{(N - M)}\sum_{i=1}^{N}\left[\dfrac{v_{c,the}(R_{i}) - v_{c,obs}(R_{i})}{\sigma_{i}}\right]^2.
\end{equation}
The sum runs over the observational data points, with $N$, the total number of data points from the observed rotation curve and $M = 3$, which is the number of free parameters.
In Eq. (\ref{chy}), $ v_{c,obs} (R_{i})$ is the observed rotational velocity related with $i$-th data point at radius $R_{i}$, $\sigma_{i}$ is the observational error bar related with each data point, and $v_{c,the}$ is the theoretical rotational velocity of approaching and receding sides extracted from the LOS velocity for an $m = 1$ perturbation in the potential model. \\
In contrast to \cite{kep}, we are able to fit the \object{WLM} rotation curve simultaneously for both approaching and receding sides with a lopsided halo potential model. Our results for value of the best-fit parameters for both of receding and approaching sides are $\epsilon_{lop} = 0.11$, $\phi_{1} = 0 \degr$, and $v_{e} = 44 (km/s)$.

%The best fitting parameters are listed in Table (\ref{bestfit}).
%Our main results for best-fit solutions for both of approaching and receding sides are summarized in table (\ref{bestfit}) which shows the constrained parameters $\epsilon_{lop}$, $v_{e} $ and $\phi_{1}$ which the rotation curve is fitted through.
 %According to the relation between $\epsilon_{iso}$ and $\epsilon_{lop}$ presented in equation  (\ref{epsilons}):
%\begin{equation}
%\epsilon_{iso}= 4  \epsilon_{lop} \left(1+\dfrac{R_{exp}}{2R}\right).
%\end{equation}

%\begin{table*}
%\centering
%\caption{Best-fit solutions for the rotation curve with an $m = 1$ perturbation in the halo potential model.}
%\label{bestfit}
 %\begin{threeparttable}
% \scalebox{1.1}{
%\begin{tabular}{ccccc}
%\hline
%& Parameters && $\chi_{r}^{2}$ for the Best-fit & \\
%\hline
 % & $\epsilon_{lop} = 0.11$, $\phi_{1} = 0 \degr $ and $v_{e} = 44 (km/s)$ && 3.8 &\\
%\hline
 %\end{tabular}}
  %\end{threeparttable}
%\end{table*}

%\begin{table}
%\centering
%\caption{Best-fit solutions for the rotation curve with an $m = 1$ perturbation in the halo potential model.}
%\label{bestfit}
%\begin{tabular}{cccccccc}
%\hline
 %&&& Parameters && $\chi_{r}^{2}$ for the Best-fit && \\
%\hline
%Approaching side & && $\epsilon_{lop} = 0.11$, $\phi_{1} = 0 \degr $ and $v_{e} = 44 (km/s)$ && 3.8 &&\\
%\hline
%Receding side & &&$\epsilon_{lop} = 0.11$, $\phi_{1} = 0 \degr $ and $v_{e} = 44 (km/s)$  && 3.8 &&\\
 %\hline
 %\end{tabular}
%\end{table} 

\begin{figure}
        \includegraphics[width=\columnwidth]{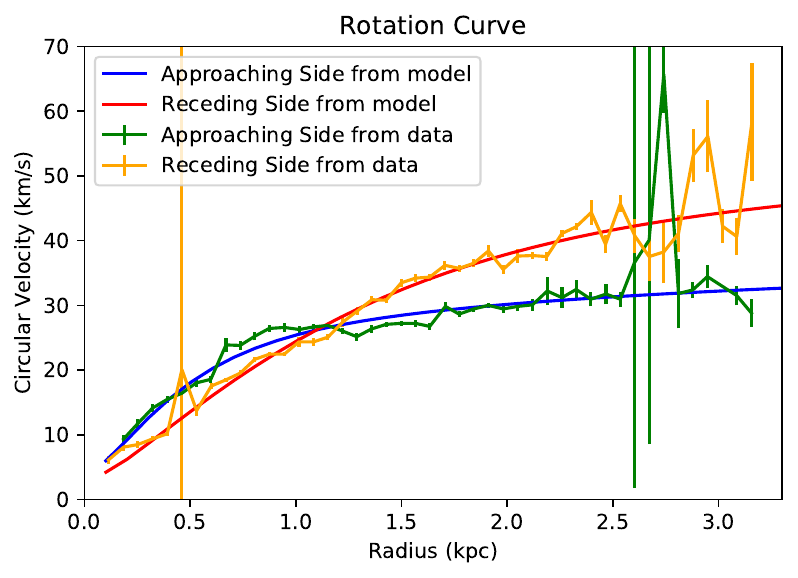}
    \caption{Asymmetric rotation curve extracted from the perturbed halo potential (lopsided halo potential) model and from the observational data \citep{kep} for the approaching and the receding sides of \object{WLM}. A perturbed halo potential (a global m=1 mode in the potential) can create kinematical asymmetry between two halves of galaxy and. By comparing the shape of rotation curve and asymmetry between approaching side and receding side obtained from lopsided halo potential model with the shape of observed rotation curves for two sides of this galaxy, we can see the good agreement between corresponding gas kinematics of a perturbed halo potential and the observed gas kinematics. }
    \label{ARC}
\end{figure}

Figure \ref{ARC} shows the extracted asymmetric rotation curve of \object{WLM} from this model associated with an $m = 1$ perturbation in the halo potential (a lopsided halo potential model) and also asymmetric rotation curve from the observational data by \cite{kep}. By comparing the shape of rotation curve and asymmetry between approaching side and receding side obtained from lopsided halo potential model with the shape of observed rotation curves for two sides of this galaxy (see Fig. \ref{ARC}), we can see the good agreement between corresponding gas kinematics of a perturbed halo potential and the observed gas kinematics. Thus, a lopsided halo potential model is able to explain the asymmetry in the kinematic data reasonably well.

\begin{figure*} 
 \subfigure[]{\includegraphics[width=\columnwidth]{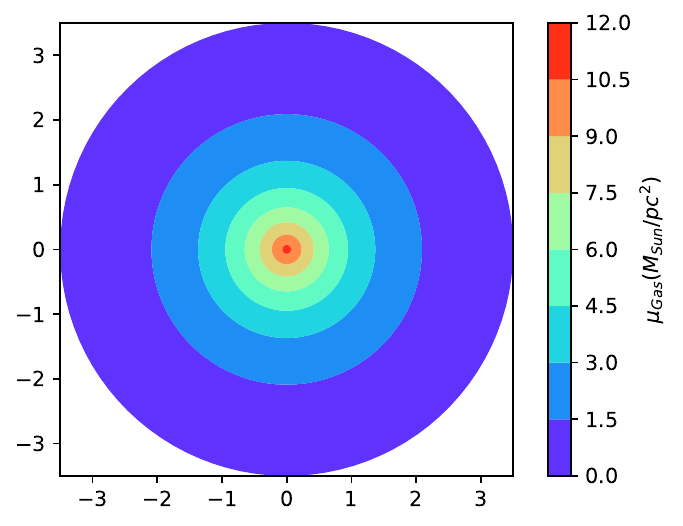} \label{non-perturbedSurfaceg}}\quad
\subfigure[]{\includegraphics[width=\columnwidth]{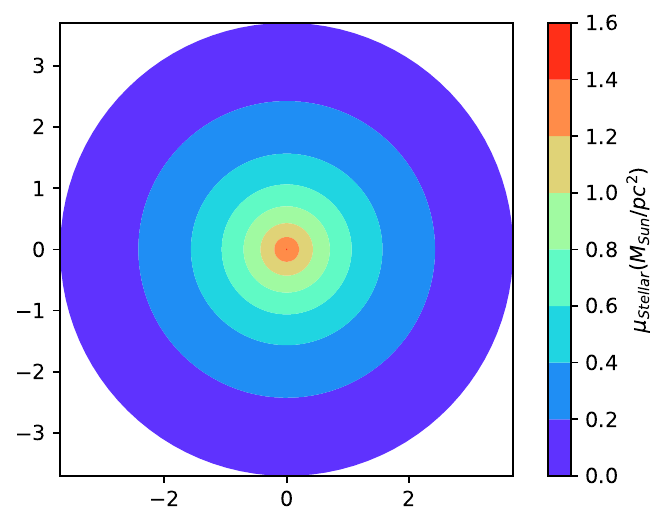} \label{non-perturbedSurfaces}}\quad\\
 \subfigure[]{\includegraphics[width=\columnwidth]{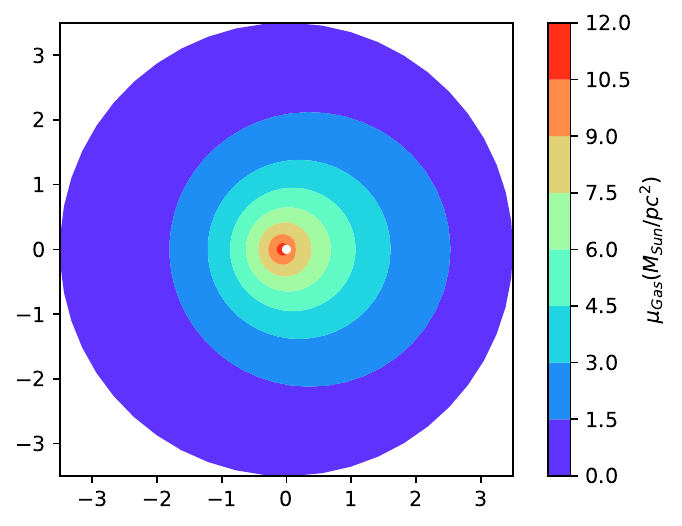} \label{figSDMg}}\quad
\subfigure[]{\includegraphics[width=\columnwidth]{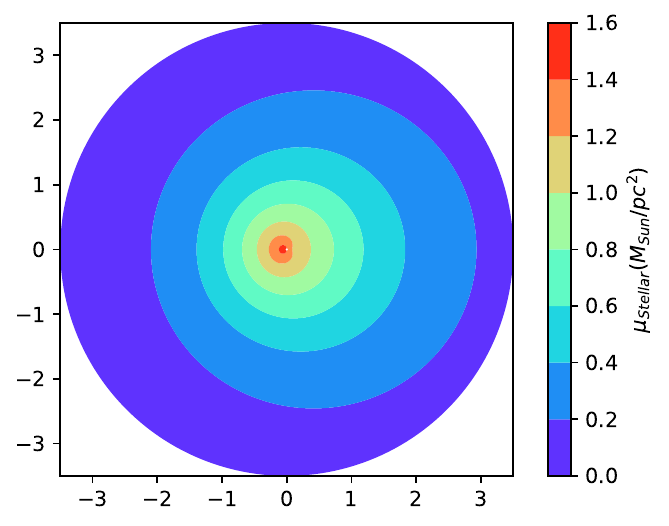}\label{figSDMs} }
\caption{Exponential surface density maps of the gaseous disk and the stellar disk in a non-perturbed halo potential and a perturbed halo potential, projected in the X-Y plane. (a) Exponential surface density map of the gaseous disk for a non-perturbed halo potential. (b)  Exponential surface density map of the stellar disk for a non-perturbed halo potential. (c) Asymmetric gas surface density map associated with a lopsided (perturbed) halo potential model-cube (contours of constant surface density for an $m=1$ perturbation in the potential). (d) Asymmetric stellar surface density map associated with a lopsided (perturbed) halo potential model cube. 
The exponential gaseous disk scale length is $R_{exp, gas} = 1.04 kpc$ and the central surface density for the total gas ($HI$ and $He$) disk is $\mu_{0, gas} = 11.2 M_{Sun}/pc^{2} $.  The exponential stellar disk scale length $R_{exp, stellar} = 1.24 kpc$ and the central surface density for stellar disk, $\mu_{0, stellar} = 1.4 M_{Sun}/pc^{2} $. 
There is symmetry between two halves of gaseous disk surface density map in Figure (a) and stellar disk surface density map in Figure (b) in a non-perturbed halo potential and there is asymmetry between two halves of gaseous disk surface density map in Figure (c) and stellar disk surface density map in Figure (d) in a perturbed (lopsided) halo potential.}
%\label{}
\end{figure*}

By considering that the non-perturbed surface density have an exponential dependence on radius as:
\begin{equation}
\mu_{un}(R)=\mu_{0}\exp\left[-\dfrac{R}{R_{exp}}\right]
,\end{equation}
where $\mu_{0}$ is the central surface density and $R_{exp}$ is the exponential disk scale length. We present the non-perturbed surface density map for the exponential gaseous disk in Fig. \ref{non-perturbedSurfaceg} and the exponential stellar disk in Fig. \ref{non-perturbedSurfaces}. \\

A lopsided gravitational halo potential can perturb an axisymmetric galactic disk as the disk surface density responds to the total asymmetry \citep{Jog2000}. Since the circular velocity varies along the perturbed orbit, the related surface density also varies as a function of the azimuthal angle $\phi$. This changes for particles (gas and stars) on these orbits are governed by the equation of continuity:
\begin{equation}
\dfrac{\partial}{\partial R}\left[ R\mu(R,\phi)v_{R}(\phi)\right] + \dfrac{\partial}{\partial \phi}\left[ \mu(R,\phi)v_{\phi}(\phi)\right] = 0
.\end{equation}
According to the equation of continuity and paper \citep{jog}, the effective surface density for an exponential disk with perturbed orbits in a lopsided halo potential may be written as:
\begin{equation}\label{EffectiveSD}
\mu(R,\phi)=\mu_{0}\exp\left[-\dfrac{R}{R_{exp}}(1-\dfrac{\epsilon_{iso}\cos\phi}{2})\right]
,\end{equation}
where $\epsilon_{iso}$ is the ellipticity of an isophote at $R$ for $m = 1$ perturbation, as follows:
\begin{equation}\label{Elliptisity}
\epsilon_{iso} =1-\dfrac{R_{min}}{R_{max}}=\dfrac{2\Delta R}{R},
\end{equation}
where $R_{max}$ is the maximum extent of an isophote and $R_{min}$ is the minimum extent of that isophote. \\
By using the equation of continuity and the relations for the orbits (coordinates and velocities), we can obtain the relation between $\epsilon_{iso}$ and $\epsilon_{lop}$ at a given radius $R$ \citep{Jog2000}:
\begin{equation}\label{epsilons}
\dfrac{\epsilon_{iso}}{\epsilon_{lop}}=4\left(1 - \dfrac{R_{exp}}{2R}\right).
\end{equation}
For this lopsided halo potential model (a global $m = 1$ in the potential) By using the Eqs. (\ref{EffectiveSD}) and (\ref {epsilons}) and programming in Python, we extracted the surface density distribution map in the X-Y plane for the gaseous disk in Fig. \ref{figSDMg} and the stellar disk in Fig. \ref{figSDMs} for an $m = 1$ perturbation in the potential with $\epsilon_{lop} = 0.11$ which is the best-fit value of this parameter for best fit to the observed rotation curve in Fig. \ref{ARC}.
These figures, which show the surface density contours of the lopsided disk (a global $m = 1$ mode), indicate the azimuthal variation in the surface density for an exponential disk in a lopsided halo potential. 
The isophotal shapes have aligned egg-shaped contours and the center of the inner isophotes are displaced from the center of the outer isophotes (see \citealt{jog}). 
The outward contours are more and more deviated from the unperturbed surface density contours, showing a more lopsided distribution in the outer regions.

\begin{figure}[ht]
\centering
\stackengine{-0.36pt}
{\kern115.3pt\includegraphics[height=2.935in, width = 2.1in]{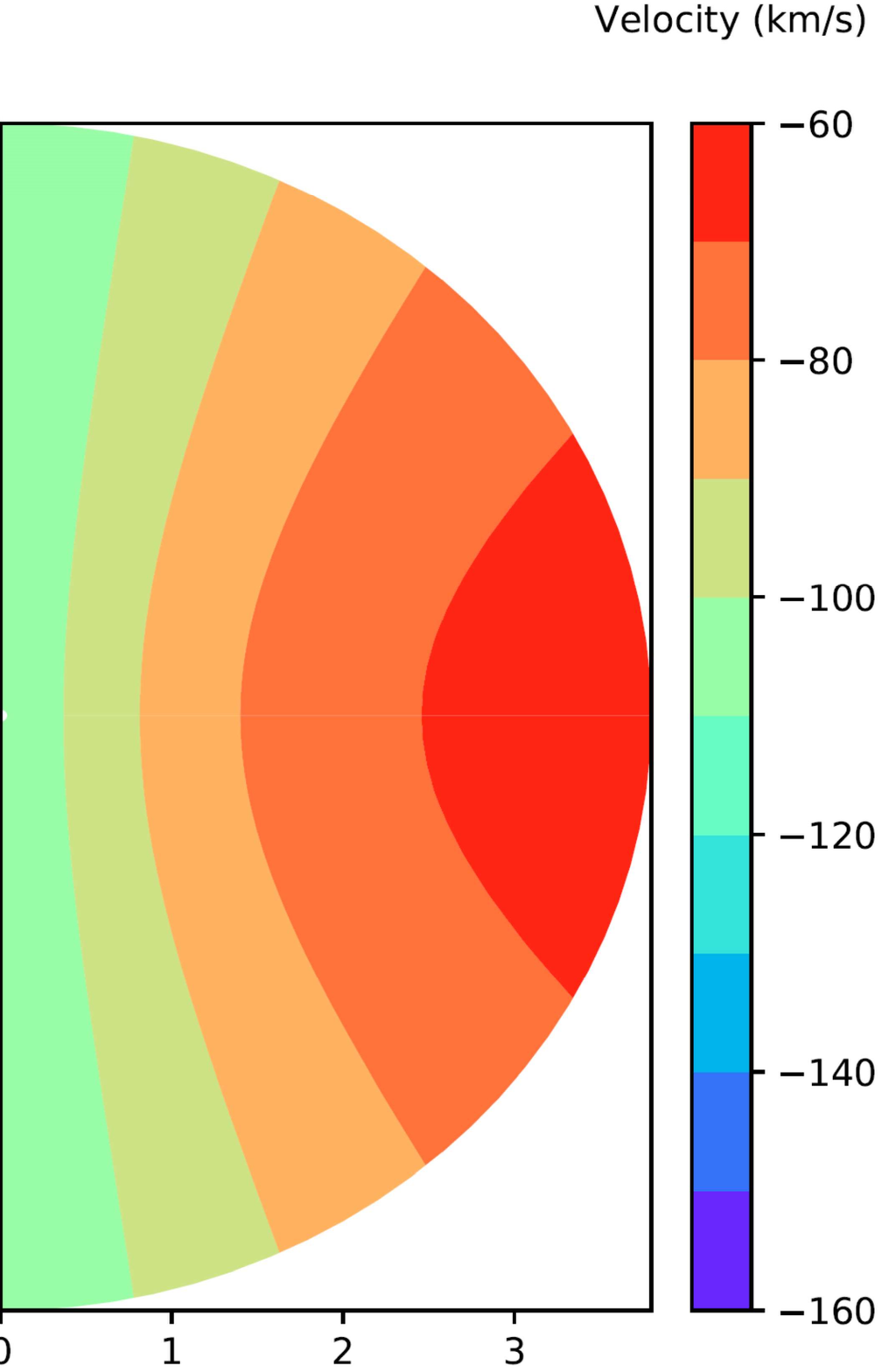}}
{\includegraphics[height=2.7in, width = 1.6in]{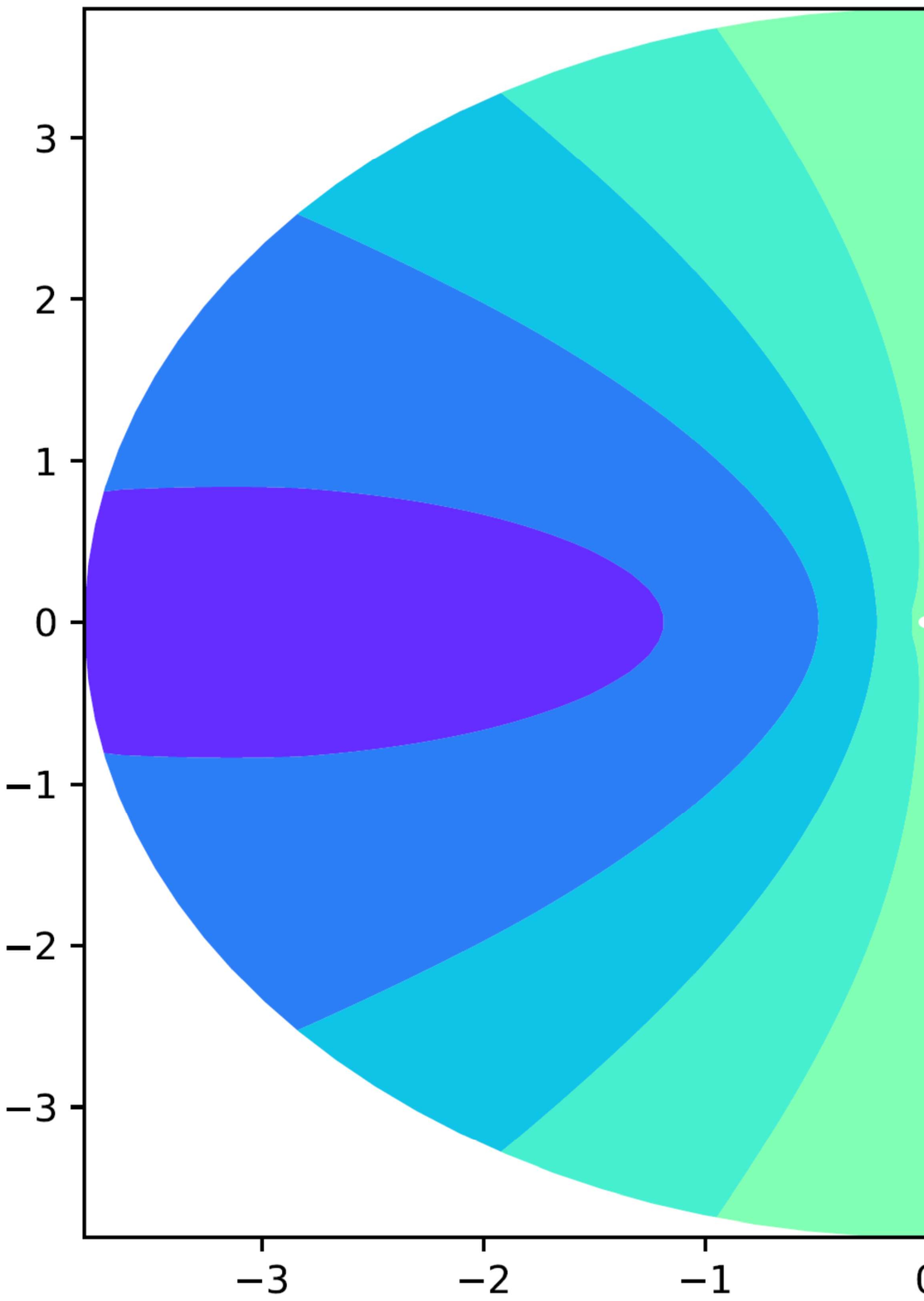}}
{O}{l}{F}{F}{L}
\caption{Perturbed velocity field map caused by a lopsided halo potential model cube with a perturbation (a global m=1 mode in the potential), projected in the X-Y plane. The perturbed velocity field map is clearly asymmetric. This model shows strong curvature of the iso-velocity contours on one side and weaker curvature on the other side. The HI kinematics is slightly lopsided. A perturbed potential can produce the asymmetric velocity fields.}
\label{figVFM}
\end{figure}

For this perturbed potential, by using the measured harmonics and other equations in this section and programming in Python, we also mapped the velocity field in Fig. \ref{figVFM} for an $m = 1$ perturbation in the potential. This perturbed velocity field map like the observational velocity field map (see Figure 5 of \citealt{kep}) is clearly asymmetric and shows the strong curvature of the iso-velocity contours on one side and the weaker curvature on the other side (iso-velocity contours are more curved on one side than them on the other side). Thus, a lopsided gravitational potential of a dark matter halo can produce an asymmetry in the rotation curve of \object{WLM}, in the velocity field map, and in the morphology of the galaxy, as well as in the gas surface density and stellar surface density (see also \citealt{jog})  between two sides of this galaxy. So such a ($\cos\phi$) asymmetry of type $m = 1$ potential perturbation  in the halo potential ( the lopsidedness of the dark matter halo) can create such a kinematical asymmetry between two sides of this galaxy.

In Table (\ref{tab2}), col. 1 gives the exponential stellar disk scale length for inner part, col. 2 gives the exponential stellar disk scale length for outer part, and col. 3 gives the exponential gaseous disk scale length. Column. 4 gives the logarithm of central surface density for stellar disk, col. 5 gives the central surface density for $HI$ gas disk, col. 6  gives the central surface density for the total gas ($HI$ and $He$) disk, and,  finally, col. 7 gives the $HI$ radius. 

\begin{table*}
\centering
\caption{The exponential disk scale lengths $R_{exp}$ and the central surface densities $\mu_{0}$ for \object{WLM}}
\label{tab2}
 \begin{threeparttable}
\begin{tabular}{ccccccc}
\hline
(1)&(2)&(3)&(4)&(5)&(6)&(7)\\
$R_{exp, \star}$ & $R_{exp, \star}$& $R_{exp, gas}$ &$ \log(\mu_{0, \star}) $ & $\mu_{0, HI}$&$\mu_{0, gas}$ & $R_{HI}$ \\
(inner) (kpc)& (outer) (kpc) & (kpc)&$\log(M_{Sun}/pc^{2})$ & $(M_{Sun}/pc^{2})$&$(M_{Sun}/pc^{2})$ &(kpc)\\
 \hline
  1.24\tnote{a}& 0.57\tnote{a} & 1.04\tnote{b} & 0.15\tnote{a} & 8 \tnote{c, d} & 11.2  & 3.5 \tnote{d} \\
  \hline
 \end{tabular}
   \begin{tablenotes}
   \item\label{3} References: (a) \citep{Hunter2012}, (b) \citep{Read2019}, (c) \citep{Oh2015} and (d) \citep{kep},  
 \end{tablenotes}
  \end{threeparttable}
\end{table*}

%\begin{table*}
%\caption{The exponential disk scale lengths $R_{exp}$ and the central surface densities $\mu_{0}$ for WLM}
%\label{tab2}
% \begin{threeparttable}
%\begin{tabular}{ccccccccc}
%\hline
%(1)&(2)&(3)&(4)&(5)&(6)&(7)&(8)&(9)\\
%$R_{exp, \star}$ & $R_{exp, \star}$& $R_{exp, gas}$ &$ \log(\mu_{0, \star}) $ &$ \log(\mu_{break, \star}) $&$\mu_{0, HI}$&$\mu_{0, gas}$ & $R_{break}$ & $R_{HI}$ \\
%(inner) (kpc)& (outer) (kpc) & (kpc)&$\log(M_{Sun}/pc^{2})$&$\log(M_{Sun}/pc^{2})$&$(M_{Sun}/pc^{2})$&$(M_{Sun}/pc^{2})$ & (kpc) &(kpc)\\
% \hline
 % 1.24\tnote{a}& 0.57\tnote{a} & 1.04\tnote{b} & 0.15\tnote{a} & -0.08\tnote{a} & 8 \tnote{c, d} & 11.2 & 1.34\tnote{a} & 3.5 %\tnote{d} \\
  %\hline
% \end{tabular}
  % \begin{tablenotes}
%   \item\label{3} References: (a) \citep{Hunter2012}, (b) \citep{Read2019}, (c) \citep{Oh2015} and (d) \citep{kep},  
% \end{tablenotes}
%  \end{threeparttable}
%\end{table*}
    
\section{Classification based on pure kinematics}\label{Classification}
Extremely lopsided mass distribution seems to occur in strongly disturbed galaxies with ongoing merger events or those that have undergone a recent merger \citep{jog}.
Such non-axisymmetry in the luminosity distribution and in the mass distribution with the isophotes which are off-centered with respect to each other is seen in the centers of mergers of galaxies \citep{Jog2006}. For an $m=1$ perturbation in the potential, the isophotal shapes have aligned egg-shaped contours \citep{jog}, as we have plotted in Figs. \ref{figSDMg} and \ref{figSDMs}, the center of the inner isophotes are displaced from the center of the outer isophotes.\\
Lopsidedness is a deviation from the ideal case that might occur in a disk. We consider what the origin of the dark matter halo lopsidedness might be for an isolated galaxy$.$  
A variety of processes such as gas accretion, interactions, and minor mergers can excite asymmetries and an $m=1$ lopsidedness in disk galaxies (e.g., \citealt{Bournaud2005, Zaritsky1997, Mapelli2008}). Thus, a disturbed structure in a galaxy may result from such interacting processes with former companions that occurred sufficiently long ago, no longer existing \citep{Fulmer2017}.
 It is thus possible that even an isolated galaxy could hold signs of an earlier minor merger such as central asymmetry and a tidal stream for a few Gyr after the merger, as shown by observational evidence for the very isolated spiral galaxy $NGC 5523, $  supporting the idea that some galaxies may have become isolated because they have experienced a historic merger with former companions \citep{Fulmer2017}. \cite{Fulmer2017} have favored a historic (in the recent past) merger as the source of perturbation that produced a long-lived asymmetry in $NGC 5523$.
Asymmetry in galaxies has  also been found in low-density environments without any direct signs of recent interactions (\citealt{Matthews1997, Matthews2002}).
Recently, \cite{Ghosh2021} carried out a simulation study of lopsided asymmetry generated in a minor merger to investigate the dynamical effect of the minor merger of galaxies on the excitation of lopsidedness. They have shown that a minor merger can trigger an $m=1$ lopsided distortions in the stellar and gas velocity field of the host galaxy and its stellar density.
We aim to determine the strength of such a deviation of the velocity field from the ideal rotating disk case and investigate whether merger could have created such a lopsidedness in the halo potential.\\

\subsection{Kinemetric analysis, quantifying asymmetries with kinemetry and merger-disk classifications based on kinematic asymmetries} 
\begin{figure}
\centering 
 \subfigure[]{ \includegraphics[width=1\columnwidth]{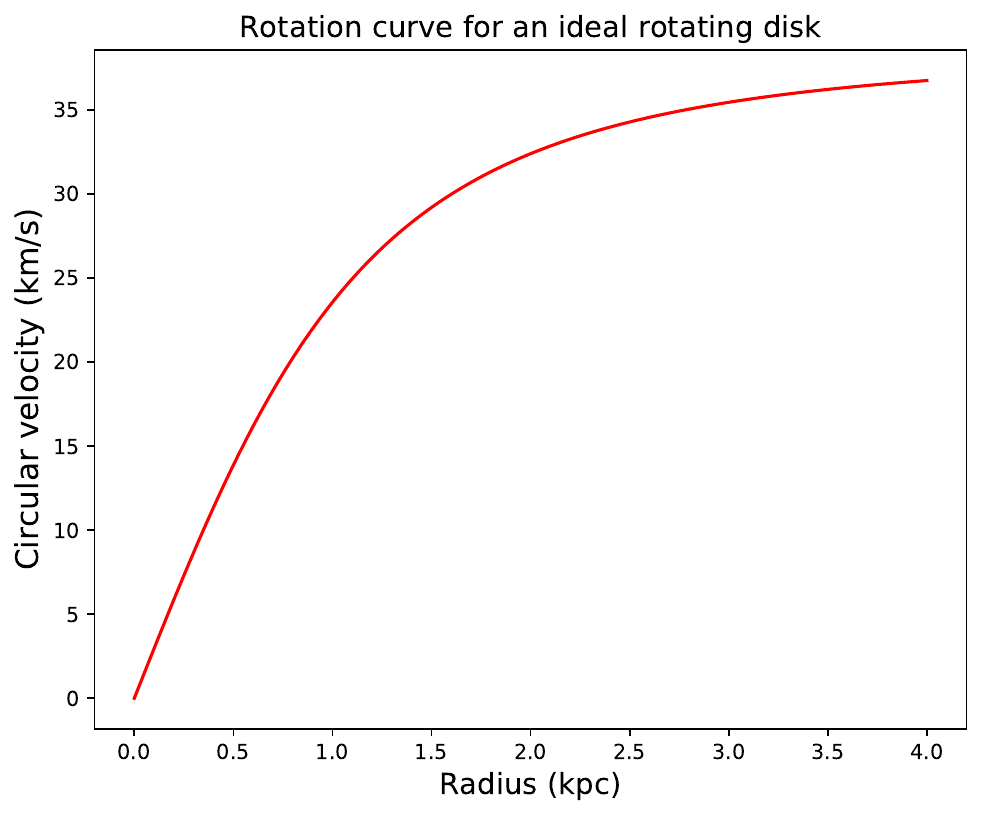} \label{c1}}\quad
\subfigure[]{ \includegraphics[width=1\columnwidth]{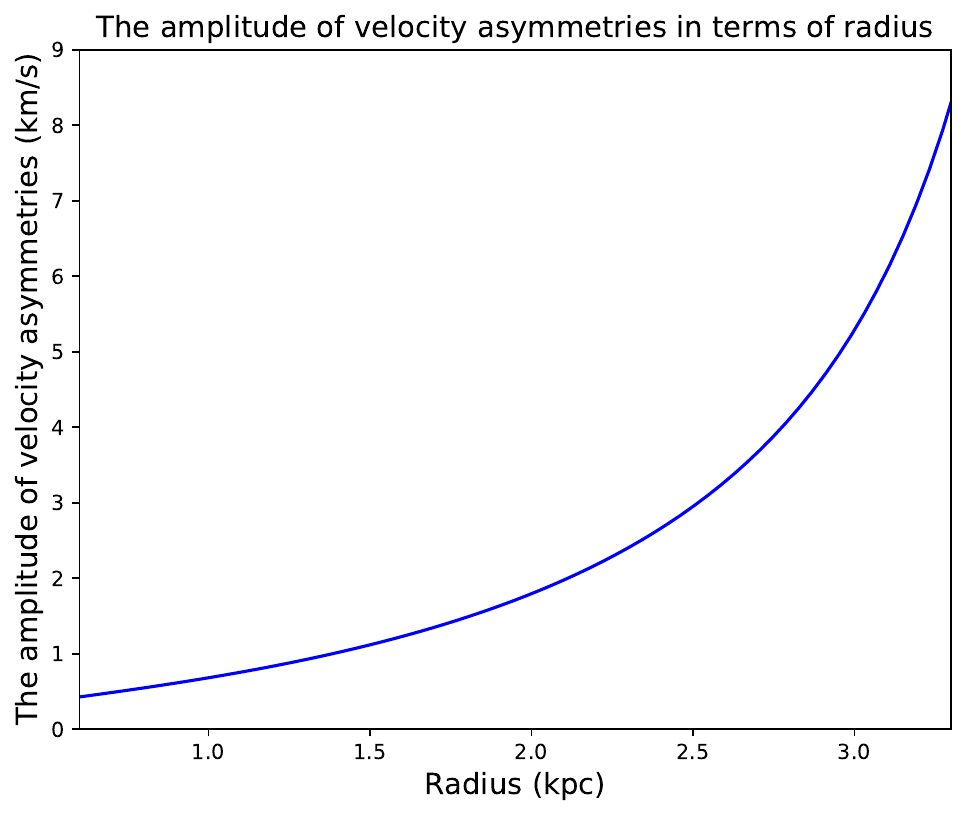} \label{k2}}\quad
\subfigure[]{ \includegraphics[width=1\columnwidth]{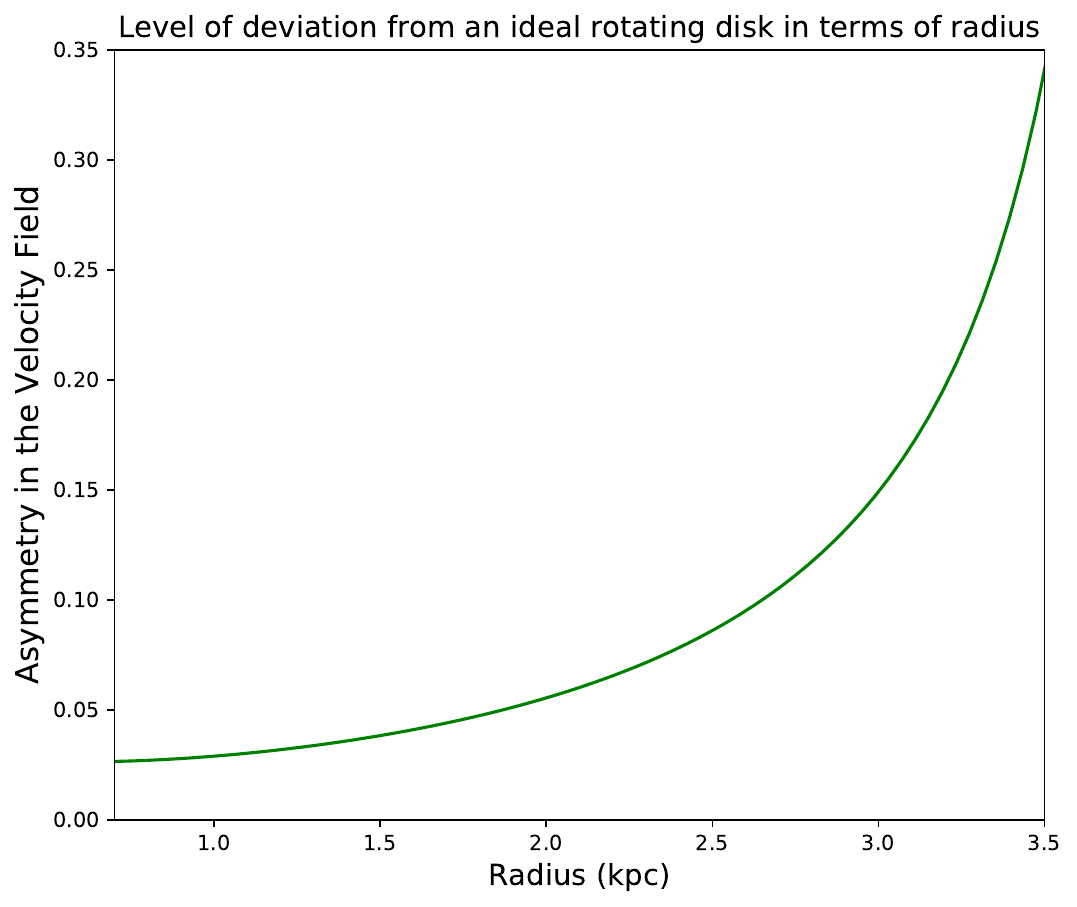} \label{vasym} }
\caption{The information about kinematic asymmetries in the velocity field of \object{WLM} and deviation of this galaxy from an ideal rotating disk. (a): Rotation curve for an ideal rotating disk in pure circular motions, $c_{1}$, in terms of radius $R$. (b): Amplitude of velocity asymmetry, $k_{2}$, in terms of radius $R$. (c): Level of deviation from an ideal rotating disk or the asymmetry in the velocity field, $v_{asym}$, in terms of radius.}
%\label{}
\end{figure}

Kinematics provides the only way to infer whether a galaxy is dominated by a relaxed rotation or by gravitational perturbation that is often linked with major merger events or minor merger and galaxy interactions. The asymmetries in the kinematic type of galaxy, such as the asymmetries that seen in the velocity field map and rotation curve of galaxy, can be measured via the kinemetry method developed and described in (\citealt{Krajnovic2006, Shapiro2008}). A velocity field map can be divided into a number of elliptical rings and along each elliptical ring, with the moment being decomposed into the Fourier series:
\begin{equation} 
v_{los} (\psi) = c_{0} + \sum\limits_{n = 1}c_{n}\cos(n\psi)+s_{n}\sin(n \psi), 
\end{equation}
where $c_{0} $, gives the systemic velocity of each elliptical ring. This yield velocity profiles to be described by a finite number of harmonic terms via Fourier series as well. So we can express the LOS velocity as:
\begin{equation} 
v_{los} (R, \psi) = v_{sys} + \sum\limits_{n = 1}c_{n}\cos(n\psi)+s_{n}\sin(n \psi).
\end{equation}
The observed LOS radial velocity fields for an ideal rotating disk (pure rotating disk in pure circular rotation) are expected to be dominated by the $\cos(\psi)$ term and are fitted with no radial or out of plane motion. So, the power in the $c_{1}$ term therefore represents the circular rotation at each ring, while power in the other coefficients (normalized to the rotation curve, $c_{1}$) represents deviations from pure circular motion \citep{Shapiro2008}. From the kinemetry decomposition, the amplitude $k_{n}$ of each Fourier harmonic order $n$ is obtained from the $c_{n}$ and $s_{n}$ coefficients \citep{Krajnovic2006}:
\begin{equation} 
k_{n} = \sqrt{c_{n}^{2} + s_{n}^{2}}.
\end{equation}
The information about kinematic asymmetries in the velocity field of galaxies is contained in higher-order terms than $k_{1}$. So, the average amplitude of velocity asymmetries $k_{avg}$ has been defined by a limited number of terms:
\begin{equation}
k_{avg} =\dfrac{1 }{m - 1} \sum\limits_{n = 2} ^{m} k_{n}.
\end{equation}
If the potential includes a perturbation of harmonic number $m$, the LOS velocity field includes $m + 1$ and $m - 1$ terms. In the case of a harmonic number $m = 1$ term in the potential, that is, a global $m = 1$ mode in the potential,  the LOS velocity field is expected to include an $m = 0$ term and an $m = 2$ term.  Therefore, for this perturbed potential model, the information on kinematic asymmetries is contained in term $k_{2}$ which is of a higher order than $k_{1}$. The amplitude $k_{2}$ of Fourier harmonic order $n = 2$ is calculated from the $c_{2}$ and $s_{2}$ coefficients:
\begin{equation} 
k_{2} = \sqrt{c_{2}^{2} + s_{2}^{2}},
\end{equation}
whereby the $c_{2}$ and $s_{2}$ coefficients are calculated in Sect. \ref{simplifying}.
By normalizing this average deviation $ k_{avg} $ to the pure circular motions, as measured by $c_{1}$, we can derive the relative level of deviation of the velocity field from that of an ideal rotating disks or asymmetry $v_{asym}$ \citep{Shapiro2008}:
\begin{equation}
v_{asym} =\left \langle \dfrac{k_{avg, v} }{c_{1}} \right\rangle_{R},
\end{equation}
where the average is over all radii. This can be applied to velocity dispersion field too. But for the velocity dispersion field,  $c_{0, \sigma}$ is the only nonzero kinemetry coefficient for an ideal rotating disk. So, the information about kinematic asymmetries in the velocity dispersion field is contained in the amplitudes $k_{1, \sigma} - k_{m, \sigma}$.  Thus, the average amplitude of velocity dispersion asymmetries $k_{avg, \sigma}$ has been defined via a limited number of terms:
\begin{equation}
k_{avg, \sigma} = \dfrac{1}{m}\sum\limits_{n = 1} ^{m} k_{n, \sigma},
\end{equation} 
and the asymmetry in the velocity dispersion field is defined as \citep{Shapiro2008}:
\begin{equation}
\sigma_{asym} = \left\langle \dfrac{k_{avg, \sigma} }{c_{1}} \right\rangle_{R}. 
\end{equation}
The average amplitude of velocity asymmetries $k_{avg}$ for the perturbed potential model (an $m=1$ perturbation in the potential) is:
\begin{equation}
k_{avg} = k_{2} 
.\end{equation}
Thus, the asymmetry $v_{asym}$ (or the relative level of deviation of the velocity field from that of an ideal disk for this model -- a global $m=1$ mode in the potential) is given by
\begin{equation}
v_{asym} = \left\langle \dfrac{k_{avg, v} }{c_{1}} \right\rangle _{R}=\left\langle \dfrac{k_{2} }{c_{1}} \right\rangle _{R}.
\end{equation}
The merger-disk classifications based on kinemetry analysis are used to kinematically classify galaxies in disk and merger.
This method relies on velocity asymmetries and can be used to distinguish unvirialized systems or those involved in major merger events from galaxies dominated by ordered rotational motion. Thus, we can differentiate disks and mergers based on symmetries of warm gas kinematics. 
We determine the strength of such a deviation of the velocity field from the ideal rotating disk case to investigate whether merger could have created such a lopsidedness in the halo potential.
An ideal rotating disk in equilibrium is expected to have an ordered velocity field, described by the spider-like diagram, and a centrally peaked velocity dispersion field \citep{Shapiro2008}. It has been shown that the merger templates have $v_{asym}$ and $\sigma_{asym}$ larger than $0.5$ \citep{Shapiro2008}. There is a defined parameter, namely, $K_{asym} = \sqrt{(v_{asym}^{2}) + (\sigma_{asym}^{2}) }$, total kinematic asymmetry, for which  $K_{asym, opt} = 0.5$ has  been established as an optimal empirical limit for separating the two classes and the classification of a system as a disk or merger can be done by this limit and major mergers or unvirialized systems can be identified via $K_{asym} > 0.5$ \citep{Shapiro2008}. \cite{Bellocchi2016} has kinematically classified galaxies as disks with a $K_{asym}  < 0.16 (0.14)$, while a merger would have a $K_{asym} > 0.94 (0.66),$ with the galaxies lying in the transition region, where disks and mergers coexist, with $0.16 (0.14) < K_{asym} < 0.94 (0.66)$.
Additionally, for the ideal case, we have ($v_{asym} \equiv 0$ and $\sigma_{asym} \equiv 0$). However, this requires a finer adjustment of the limit distinguishing mergers from the rotating disk.\\
For \object{WLM}, in Fig. \ref{c1}, we plot $c_{1}$,which is the pure circular velocity in terms of radius namely, the rotation curve for an ideal rotating disk in pure circular motions. To obtain the deviations from pure rotational velocity (circular motion), we plot the amplitude of velocity asymmetries, $k_2$, in terms of radius for an $m = 1$ perturbation in the potential in Fig. \ref{k2} based on kinematic asymmetries. In addition, we have determined the strength of deviation of the velocity field, $v_{asym}$, from the ideal rotating disk case (a purely rotating disk) by obtaining the relative level of deviation of the velocity field from that of an ideal rotating disks or the asymmetry in the velocity field, $v_{asym}$,  for the perturbed potential model from an ideal disk (with $v_{asym} = 0$ and $\sigma_{asym} = 0$ ) in terms of radius in Fig. \ref{vasym} by normalizing the average deviation $k_{avg} $ (here $k_{avg} = k_2$) to the pure circular motions, as measured by $c_{1}$.\\
To obtain Figs. \ref{c1}, \ref{k2}, and \ref{vasym}, we averaged between two sides (approaching and receding sides), so we neglected the asymmetry between the two sides, which is the dominant asymmetry in the \object{WLM} velocity field. Also, in addition to $v_{asym}$, $K_{asym}$ contains $\sigma_{asym}$ too, and $\sigma_{asym}$ is greater than zero, since the dispersion map indeed presents a minimum at the center, which is not expected for an ideal rotating disk \citep{2017bookHammer}. Thus, $K_{asym}$ value is expected to be higher than $v_{asym}$, which is shown in Fig. \ref{vasym}. 
Then it is possible that \object{WLM} lies in the transition region, in which disk and merger coexist. So, a merger may be one of the possible origins of the dark matter halo lopsidedness for this isolated galaxy (according to the classification by \citealt{Bellocchi2016}).

\subsection{Classification based on $\sigma$ centering} 
There is a parameter $\Delta r$  representing the distance between the $\sigma$ peak ($c_{\sigma}$) and dynamical centre ($c_{VF}$) \citep{Flores2006}:
\begin{equation}
\Delta r = \vert c_{\sigma} - c_{VF} \vert.
\end{equation}
The kinematics of IMAGES galaxies has been visually classified using this parameter and and the alignment of both kinematic and optical P. A (\citealt{Flores2006, Yang2008}). It has lead to three classes of kinematics, described as follows.

The first are rotating disks, which have a prominent signature in their velociy dispersion map, expressed as a central peak at the dynamical centre (with $\Delta r \sim 0$). The velocity field shows an ordered gradient and the rotation axis in the velocity field almost aligned with the optical major axis (the kinematical major axis is aligned with the morphological major axis) and the $\sigma$-map (velocity dispersion map) shows a single peak close to the kinematical center, $\sigma$ maps should show a clear peak near the galaxy dynamical centre where the velocity gradient in the rotation curve is the steepest (\citealt{Flores2006, Puech2006, Yang2008}).

Next, in perturbed rotations, the velocity field shows rotation (the kinematics shows all the features of a rotating disk) and the axis of rotation in the velocity field is aligned with the optical major axis, but the $\sigma$-map shows a peak that clearly shifted away from the kinematical centre (with $\Delta r > 0$) or does not show any peak (\citealt{Flores2006, Puech2006, Yang2008}).

Finally, complex kinematics exist in systems with a near-chaotic velocity field, while both velocity field map and $\sigma$-map (velocity dispersion map) display complex distributions; this is very different from what is expected for rotating disks. Here, neither the velocity field map nor the $\sigma$-map (velocity dispersion map) are compatible with regular rotation, including the velocity fields that are not aligned with the optical major axis (\citealt{Flores2006, Puech2006, Yang2008}, see also \citealt{Hung2015}). \\
In a minor merger, the disk is not destroyed and the kinematics of the remnant does not  appear too complex. The galaxy should be observed to still be rotating along its main optical axis. with a dispersion map that does not show any peak at the centre. This could correspond to  the perturbed galaxies \citep{Puech2006}.\\
An ideal rotating disk in equilibrium is: expected to have an ordered velocity field that is described by the spider diagram structure and a centrally peaked velocity dispersion field, with the observed velocity fields fit with circular velocity and with no radial or out of plane motions (non$-$circular motions) \citep{Shapiro2008}.
The galaxies with the main kinematics gradient well aligned with respect to the principal morphological axis can be described as a rotating disk \citep{plummer191}.

For \object{WLM}, the rotation (an ordered velocity field, described by the iso-velocity contours with a spider diagram structure) is seen in the observational velocity field map, with the velocity field showing rotation (see Figure. 5 of \citealt{kep}). The kinematics shows all the features of a rotating disk and the axis of rotation in the velocity field follows the optical major axis (see optical images of \object{WLM} in Little Things Data\footnote[1]{\url{https://science.nrao.edu/science/surveys/littlethings/data/wlm.html}}), but the velocity dispersion (Second moment) map for \object{WLM} does not show any peak at the center, which would otherwise be expected for an ideal rotating disk (see Figure. 6 of \citealt{kep} and Figure. 6, bottom left of \citealt{Rog2020}). The strong asymmetry between two sides (approaching and receding sides) of the galactic rotation curve indicates an abnormal rotation. Thus, this galaxy is found not to be an ideal rotating disk, but a rotating disk with perturbed rotations and non-circular motions, in addition to pure circular motion, and its kinematics is classified as a perturbed rotation.\\
  
\subsection{Morpho-kinematic classification}
\cite{Neichel2008} found a fine agreement between the kinematic classification and classification based on pure morphological analysis. Furthermore, \cite{Hammer2009} defined three different morpho-kinematical classes: r1) rotating spiral disks are galaxies that show a rotating velocity field and a dispersion peak at the dynamical centre
(see e.g., \citealt{Flores2006}) and with the appearance of a spiral galaxy; 2) non-relaxed (Nonvirialized) systems are galaxies with velocity fields that are incompatible from a rotational velocity field and are characterized by a peculiar morphology; 3) semi-relaxed (Semivirialized) systems show either a rotational velocity field and a peculiar morphology or a velocity field incompatible from rotation and a spiral morphology. \\
This kind of classification and the degree of virialization scheme was previously presented by \cite{Hammer2009} and on its basis, \object{WLM} apepars to be in an intermediate category between non-virialized and semi-virialized systems.

\section{Conclusion}\label{Conclusion}
In this study, we investigate the origin of the strong asymmetric rotation curve of the dwarf irregular galaxy \object{WLM} by examining whether an $m = 1$ perturbation (lopsidedness) in a halo potential could be considered the mechanism behind such kinematical asymmetry. To do so, we fit the rotation curve of a lopsided halo potential model to that of \object{WLM}. 
There is a good agreement between corresponding gas kinematics of a perturbed halo potential and the observed gas kinematics of \object{WLM}.
In constrast to \cite{kep}, our model provides a good fit to the \object{WLM} rotation curve simultaneously for both approaching and receding sides, with a lopsided halo potential model with the best-fitting parameters of $\epsilon_{lop} = 0.11$, $\phi_{1} = 0 \degr$, and $v_{e} = 44 (km/s)$.
For this best-fit model, we mapped the surface density field for gas disk and stellar disk, as well as the velocity field. The latter shows asymmetric characteristics that is comparable to the observed one, as shown, for instance, in Figure. 5 in \citep{kep}.
We conclude that a lopsided halo potential model can explain the asymmetry in the kinematic data reasonably well.

\object{WLM} shows an ordered velocity field \citep[see Figure.~5 of][]{kep} and the kinematical axis is well aligned with the optical major axis. However, its velocity dispersion map does not show any peak at the centre (see Fig. 6 of \citealt{kep} and Fig.~6 of \citealt{Rog2020}), which is not expected for an ideal rotating disk. 
We studied the kinematical classification of the velocity field of \object{WLM} with various methods (see Sect. 4) and we found that its velocity field is significantly perturbed due to both its asymmetrical rotation curve and also  its peculiar velocity dispersion map. 
Based on a kinemetry analysis, we determined the strength of such a deviation of the velocity field from the ideal rotating disk case by obtaining the relative level of deviation of the velocity field from that of an ideal rotating disk case.
Thus, it is possible that \object{WLM} lies in the transition region, where disk and merger coexist. Thus, a merger may indeed be one of the possible origins of the dark matter halo lopsidedness for this isolated galaxy. In conclusion, it appears that the rotation curve of \object{WLM} diverges significantly from that of an ideal rotating disk, which may significantly affect investigations of its dark matter content.

\begin{acknowledgements}
The authors would like to thank the anonymous referee for her/his precise and constructive comments. M. Khademi would like to thank the GEPI Group (Observatoire de Paris) for their hospitality and their support. M. Khademi and S. Nasiri would like to thank Shahid Beheshti University.
\end{acknowledgements}

%%%%%%%%%%%%%%%%%%%%%%%%%%%%%%%%%%%%%%%%%%%%%%%%%%


 \bibliographystyle{aa} % style aa.bst
\bibliography{myrefAandA} % your references Yourfile.bib
%\bibliographystyle{mnras}
%\bibliography{myrefAandA} % if your bibtex file is called example.bib
%\footbibliography{myrefAandA}

% This method is tedious and prone to error if you have lots of 

%%%%%%%%%%%%%%%%%%%%%%%%%%%%%%%%%%%%%%%%%%%%%%%%%%

%%%%%%%%%%%%%%%%% APPENDICES %%%%%%%%%%%%%%%%%%%%%

%%%%%%%%%%%%%%%%%%%%%%%%%%%%%%%%%%%%%%%%%%%%%%%%%%

% Don't change these lines
        % typesetting comment
\label{lastpage}

\end{document}